\documentclass[prd,amsfonts,onecolumn,nofootinbib,showpacs]{revtex4}

\usepackage{graphicx}

\newcommand{\be}{\begin{equation}}
\newcommand{\ee}{\end{equation}}
\newcommand{\bea}{\begin{eqnarray}}
\newcommand{\eea}{\end{eqnarray}}

\newcommand{\gapp}{\mathrel{\raise.3ex\hbox{$>$}\mkern-14mu
              \lower0.6ex\hbox{$\sim$}}}
\newcommand{\lapp}{\mathrel{\raise.3ex\hbox{$<$}\mkern-14mu
              \lower0.6ex\hbox{$\sim$}}}


\begin{document}

\title{Flaxino dark matter and stau decay}

\author{Eung Jin Chun$^{1}$, Hang Bae Kim$^{2}$,
Kazunori Kohri$^{3}$, and David H. Lyth$^{3}$}


\affiliation{
$^ 1$ Korea Institute for Advanced Study, Seoul 130-722, Korea, \\
$^ 2$ BK21 Division of Advanced Research \& Education in Physics,
Hanyang University, Seoul 133-791, Korea, \\
$^ 3$ Physics Department, Lancaster University, Lancaster,  LA1 4YB, UK}

\begin{abstract}
If the spontaneous breaking of Peccei-Quinn symmetry comes from soft
supersymmetry breaking, the fermionic partners of the symmetry-breaking
fields have mass of order the gravitino mass, and are called flatinos. The
lightest flatino, called here the flaxino, is a CDM candidate if it is the
lightest supersymmetric particle. We here explore flaxino dark matter
assuming that the lightest ordinary supersymmetric particle is the stau,
with gravity-mediated supersymmetry breaking.
The decay of the stau to the flaxino is  fast enough not to
spoil the standard predictions of Big Bang Nucleosynthesis, and
its track and decay can be seen in future colliders.
\end{abstract}

\maketitle

\section{Introduction}

By observing the temperature anisotropy of Cosmic Microwave Background
(CMB), the WMAP Collaboration reported that the density parameter of
dark matter (DM)  or cold dark matter (CDM)  $\Omega_{\rm DM, obs}$ is
\begin{eqnarray}
    \label{eq:OmegaWMAP3}
    \Omega_{\rm DM, obs}h^{2} = 0.104^{+0.007}_{-0.010},
\end{eqnarray}
at 68 \% C.L.~\cite{Spergel:2006hy} where $h$ is the normalized Hubble
constant.  One of the most popular candidates for DM is the  lightest
supersymmetric particle (LSP) in supersymmetric theories with
R-parity.  Recent studies showed that Big Bang Nucleosynthesis (BBN)
puts interesting limitations on the candidates of LSP DM.

A long-lived next LSP (NLSP) whose lifetime is longer than $0.1$
sec is dangerous for the successful BBN because the decaying NLSP
produces a lot of (high-energy) daughter particles such as photons
and hadrons during/after BBN, which can destroy $^{4}$He (and D)
and non-thermally produce another light elements
copiously~\cite{radBBN,hadBBNold,hadBBNrec}.  In particular, a
non-thermal hadron emission due to the decaying NLSP is severely
constrained by the observational light-element abundances of D,
$^{4}$He, $^6$Li, and $^7$Li~\cite{hadBBNold,hadBBNrec}.

In the popular case of the neutralino (gaugino or higgsino) LSP, as is
well-known, the unstable gravitino decay puts stringent upper bound on
the reheat temperature after primordial inflation, which could be
problematic for various cosmological considerations, e.g., of the
inflation models. This problem can be avoided if the gravitino is the
LSP. (See also~\cite{Moroi:1993mb} for constraints on the reheat
temperature in the gravitino LSP scenario).  However, the decays of
ordinary supersymmetric particles to the gravitino turn out to be
dangerous as well.  The case of neutralino NLSP, because the branching
ratio into hadrons is close to the order of ${\cal O}(1)$, is almost
excluded~\cite{Feng:2004mt,Steffen:2006hw} by considering BBN
constraints~\cite{hadBBNold,hadBBNrec}.
Slepton NLSP scenarios appear much more favorable than neutralino NLSP
scenarios in  cosmology.  Although the sneutrino NLSP scenario should
be possible because of the smaller hadronic and radiative decaying
modes, the primordial abundance of the sneutrino is  too small to
satisfy the dark matter density for its daughter LSP
particle~\cite{Feng:2004zu,Kanzaki:2006hm}. Therefore, we would need
another production mechanism of the LSP in cosmology, which usually requires
a fine tuning.

Recently, the stau NLSP scenario has been discussed a lot as there are
specifically severer/more interesting problems in BBN related to  the
long-lived negatively charged
particle~\cite{Pospelov:2006sc,Kohri:2006cn,Kaplinghat:2006qr,Cyburt:2006uv}.
See also long-term problems discussed in literature~\cite{CargedDMold}
and  topics in doubly-charged particles~\cite{doubleC}.  If the
lifetime of the negatively charged particle $C^{-}$ such as
$\tilde{\tau}^{-}$ is longer than 10$^{2}$ sec, bound states with
ambient light elements can be produced. In particular,   the bound
state with $^{4}$He denoted by ($^{4}$He,$C^{-}$), which is produced
after 10$^{3}$ sec, significantly enhances the reaction rate
D+($^{4}$He,$C^{-}$) $\to$ $^{6}$Li + $C^{-}$ +
$\gamma$~\cite{Pospelov:2006sc} by a factor of $\sim {\cal
O}(10^{7})$~\cite{Hamaguchi:2007mp}. Then $^{6}$Li is over-produced by
the catalyzed BBN, and this scenario is severely constrained by
observations~\cite{Kawasaki:2007xb}.

\smallskip

Various attempts
\cite{Bird:2007ge,CBBNsolution,Buchmuller:2007ui,CBBNrecent} have been
made to tackle this problem and realize a successful dark matter
production mechanism.  In this paper, we suggest  the ``flaxino'' DM
as a viable option to avoid the above-mentioned difficulties for the
neutralino or slepton NLSP.  The flaxino is an axino-like fermion
appearing in flat-direction axion models where the spontaneous
breaking of Peccei-Quinn symmetry comes from soft supersymmetry
breaking and thus all of the fermionic partners of the
symmetry-breaking fields have mass of order the gravitino mass. The
flaxino has a coupling to the ordinary superparticles suppressed by
the axion scale $F_a$ and thus attractively leads to a fast decay of
NLSP,  which should be compared with the gravitino coupling suppressed
by the Planck scale.

\section{LSP from  flat-direction axion sector}

The strong CP problem can be naturally resolved by assuming the
Peccei-Quinn $U(1)$ symmetry spontaneously broken at the usual high
scale, $F_a=10^{10-12}$ GeV, and thus leads to an axion of the KSVZ
\cite{KSVZ} or DFSZ type \cite{DFSZ}. In the supersymmetric standard
model, the realization of the DFSZ axion model nicely leads to the
origin of the $\mu$ term \cite{CKN}. Furthermore, the axion scale
$F_a$ appears as the geometric mean of the Planck scale and soft
supersymmetry breaking scale by invoking an almost flat potential for
the Peccei-Quinn symmetry breaking fields \cite{MSY,CCK}.  For
instance, let us introduce two singlet fields $P,Q$ with appropriate
Peccei-Quinn charges to allow \cite{Chun:2000jr,Chun:2000jx}
\begin{equation}
    \label{eq:superpot}
 W = h {P Q\over M_P} H_1 H_2 + f { P^3 Q
 \over M_P}
\end{equation}
where $M_P$ is the reduced Planck scale. Then, the $\mu$ term and
axion scale can be related by $\mu= h \langle P \rangle \langle Q
\rangle/M_P$ and $F_a \sim \langle P \rangle, \langle Q \rangle \sim
\sqrt{\tilde{m}M_P} \sim 10^{10}$ where $\tilde{m}\sim 10^{2-3}$ GeV
is the typical soft mass scale (correctly $F_{a} = \sqrt{\langle P
\rangle^{2} + 9\langle Q \rangle^{2}}$ in this model). This is below
the scale $F_{a} \gtrsim 10^{12}$ GeV required to make the axion the
CDM. This model contains two scalar flatons $F_{1,2}$, one
pseudoscalar flaton $F'$, and two flatinos $\tilde{F}_{1,2}$. This
kind of flat DFSZ axion models lead to thermal inflation
\cite{LythStewart} and the model parameters are rather severely
constrained in order to avoid the over-production of unwanted relics
like axions or the LSP after thermal inflation \cite{CCK}. An
interesting collider implication of such models has been pointed out
in Ref.~\cite{Martin:2000eq}. The LSP can be some linear combination
of the fermionic parts of $P$ and $Q$, which we now call ``flaxino''
(denoted by $\tilde{F}_{1}$), and then the usual neutralino or stau
NLSP decays to the flaxino mainly through its mixing with higgsinos
driven by the $\mu$ term. Although the corresponding decay lengths can
be larger than the size of detectors, the copious production of the
NLSPs in future colliders may enable us to observe their decays for
the low axion scale $F_a \sim 10^{10}$ GeV \cite{Martin:2000eq}. The
general idea of the axino DM has been put forward in Ref.~\cite{Covis}
(see also a recent related paper~\cite{Brandenburg:2004du}).

\smallskip

Some of the current authors have investigated the scenario of the
``flaxino'' LSP  and the neutralino NLSP in the flat DFSZ axion models
 using gravity-mediated SUSY breaking~\cite{Chun:2000jr,Chun:2000jx}.
 The axino LSP in KSVZ type models has been
considered in Ref.~\cite{Brandenburg:2005he}.  The KSVZ type models
(hadronic axion models) do not agree with the framework of the
flat-direction models because of the overproduction of the flaxino
abundance, which is excluded by  BBN \cite{Chun:2000jr}.
Extending the analysis of Ref.~\cite{Chun:2000jx} where the
neutralino NLSP is assumed, we will consider the possibility of
the stau NLSP which is also allowed in a large parameter space of
the MSSM with the minimal supergravity scheme.

Let us first note that the DFSZ models predict a shorter lifetime of
the  stau NLSP compared to the KSVZ models since the
stau--axino/flaxino coupling arises at tree level in the former
case and at one-loop level in the latter. As we will see later,
even in the case of $F_{a} \lesssim 10^{14}$ GeV, the stau
lifetime does not exceed 10$^{3}$ sec for typical mass scales of
SUSY particles in the DFSZ axion models and does not induce the
$^{6}$Li over-production problem. This is a distinct feature of
the DFSZ axion models. Recall that the usual upper bound, $F_{a}
\lesssim 10^{12}$ GeV, can be relaxed in late-time entropy
production scenarios such as thermal inflation models which we are
considering at present.

From the viewpoint of detectabilities of a superparticle at Large
Hadron Collider (LHC), such a long-lived charged particle is promising
as it can be easily traced within the detector.  Moreover, quite
recently Refs.~\cite{Hamaguchi:2004df,Feng:2004yi,De Roeck:2005bw}
have discussed aggressive future plans to place  new stopper, e.g.,  5
- 10 meter wall made of iron,  water tanks, rocks and so on, outside
regular detectors in LHC such as ATLAS or CMS to electromagnetically
stop the produced staus. If these plans are realized, we will be able
to  measure the stau's lifetime even if its decay length is longer
than the size of the  detectors ($ \lesssim {\cal O}(10)$ m), and
reconstruct the track of the neutral LSP by catching other daughter
particles. Then, according to a similar discussion in Brandenburg et
al's analysis~\cite{Brandenburg:2005he}, observing three-body decay of
the staus will enable us to distinguish between signals of the
axino/flaxino LSP and the gravitino LSP.

We refer to  \cite{Seto:2007ym} for another topics of axino DM and
gauge-mediation supersymmetry breaking models, and
Ref.~\cite{Kawasaki:2007mk} for general cosmological constraints on a
decaying saxion or flaton. See also \cite{Choi:2008zq} for an inverted
mass case such as the axino NLSP and the neutralino LSP, and
\cite{Asaka:2000ew} for the gravitino NLSP and the axino LSP in
gauge-mediated supersymmetry breaking models.

\section{Stau Lifetime}
\label{sec:lifetime}

Following Ref.~\cite{Martin:2000eq}, we calculate the lifetime of the
stau NLSP which decays into the flaxino LSP $\tilde{F}_{1}$ in the
DFSZ axion models. For the sake of simplicity we show only mixing
between $\tilde{F_{1}}$ and the usual four neutralino components;
bino, wino, and two higgsinos. This case is sufficient if we are not
interested in the heavier flaxino $\tilde{F_{2}}$ whose mass is larger
than $m_{\tilde{N}_{1}^{0}}$. Then, we get
\begin{eqnarray}
    \label{eq:Gammastau}
    \Gamma(\tilde{\tau}_{1}\to\tau\tilde{F}_{1})
= \frac{m_{\tilde{\tau}_{1}}}{16\pi}
  \sqrt{\lambda(1,r_{0}^{2},r_{\tau}^{2})}
  \left[
    \left(
      |a_{0}^{\tilde{\tau}_{1}}|^{2}
      + |b_{0}^{\tilde{\tau}_{1}}|^{2}
    \right)
    \left(
      1 - r_{0}^{2} - r_{\tau}^{2}
    \right)
 - 4 r_{0} r_{\tau} {\rm Re}
   \left(a_{0}^{\tilde{\tau}_{1}} b_{0}^{\tilde{\tau}_{1}*} \right)
 \right]
\end{eqnarray}
where $m_{\tilde{\tau_{1}}}$ is the lightest stau mass ($\equiv
m_{\tilde{\tau}}$ in this paper), $r_{0}
=m_{\tilde{N}_{1}}/m_{\tilde{\tau}_{1}}$, $r_{\tau}
=m_{\tau}/m_{\tilde{\tau_{1}}}$ with 
\begin{eqnarray}
    \label{eq:lambda}
    \lambda(x, y, z) = x^{2}+y^{2}+z^{2}- 2 x y -2 yz - 2zx,
\end{eqnarray}
and
\begin{eqnarray}
    \label{eq:AandB1}
    a_{0}^{\tilde{\tau}_{1}} =
     \sqrt{2} s_{\tilde{f}}
    \left[
      g g T_{3f} N_{02}^{*} +g'
        \left(
          q_{f} - T_{3f} N_{01}^{*}
        \right)
    \right]
   - c_{\tilde{f}}^{*} g N_{03}^{*} m_{\tau}/\sqrt{2} c_{\beta}m_{W},
\end{eqnarray}
\begin{eqnarray}
    \label{eq:AandB2}
    b_{0}^{\tilde{\tau}_{1}} =
     \sqrt{2} c_{\tilde{f}} g' q_{f} N_{01}
    + s_{\tilde{\tau}}g  N_{03} m_{\tau}/\sqrt{2}c_{\beta}m_{W} .
\end{eqnarray}
Here $m_{\tau}$ and $m_{W}$ are the masses of the tau lepton and the
weak boson, $g$ and $g'$ are the weak and electromagnetic coupling
constants, respectively, and the Z-boson couplings to quarks and
leptons for the tau lepton are give by $(T_{3\tau},q_{\tau}) =
(-1/2,-1)$. The matrix elements $N_{0i}$ ($i$=1,2.3,4) are
flaxino--neutralino components of a unitary matrix $N_{ij}$ which is
the 5$\times$5 mass matrix of four neutralinos plus one flaxino,
$M_{kl}^{(5)}$ ($k,l$=0,1,2.3,4) in the
$(\tilde{F}_{1},\tilde{B},\tilde{W}^{0},\tilde{H}^{0}_{d},\tilde{H}^{0}_{u})$
basis. It can be diagonalized as
\begin{eqnarray}
    \label{eq:diag5}
    N_{jk}^{*}M_{kl}^{(5)}N_{lj} = \delta_{ij} m_{\tilde{N}_{i}},
\end{eqnarray}
where 
\begin{eqnarray}
    \label{eq:M5}
    M^{(5)} =
\left(
  \begin{array}{ccccc}
 m_{\tilde{F}_{1}} & 0 &  0 & - \delta_{1} s_{\beta} \mu
  & - \delta_{1} c_{\beta} \mu
  \\
 0  & M_{1} & 0  & - c_{\beta} s_{W} m_{Z} &   s_{\beta} s_{W} m_{Z}
 \\
 0 &  0 & M_{2} & c_{\beta} c_{W} m_{Z} & - s_{\beta} c_{W} m_{Z}
 \\
 - \delta_{1} s_{\beta} \mu & - c_{\beta} s_{W} m_{Z} &
  c_{\beta}c_{W}m_{Z} & 0  & -\mu
  \\
  -\delta_{1} c_{\beta} \mu & s_{\beta} s_{W} m_{Z}&
  -s_{\beta} c_{W} m_{Z} & -\mu  & 0
  \end{array}
\right)
\end{eqnarray}
Here $s_{\beta} = \sin\beta$, $c_{\beta} = \cos\beta$,  $s_{W} =
\sin\theta_{W}$, and $c_{W} = \cos\theta_{W}$  with the Weinberg angle
$\theta_{W}$, and  
\begin{eqnarray}
    \label{eq:delta1}
    \delta_{1} \equiv \frac{v}{F_{a}}
\frac{\sqrt{x^{2}+1}}{x}
\left(\cos{\tilde{\phi}}+x \sin{\tilde{\phi}} \right),
\end{eqnarray}
with the electroweak scale $v$ = 264 GeV,  $x \equiv \langle P\rangle
/ \langle Q \rangle$, the mixing angle $\tilde{\phi}$ between
$\tilde{F}_{1}$ and $\tilde{F}_{2}$ which is expressed by using $x$ to
be $\cos2\tilde{\phi} = -1/\sqrt{x^{2}+1}$ and $\sin2 \tilde{\phi} =
-x/\sqrt{x^{2}+1}$~\cite{Chun:2000jx}. The mass eigenstate labels
follow the convention: $m_{\tilde{F}_{1}} \simeq m_{\tilde{N}_{0}} <
m_{\tilde{N}_{1}} < m_{\tilde{N}_{2}} < m_{\tilde{N}_{3}} <
m_{\tilde{N}_{4}}$ (Note that $m_{\tilde{F}_{2}}$ can be larger than
$m_{\tilde{N}_{1}}$ although we did not write it down explicitly
in~(\ref{eq:M5})). In the stau sector, the diagonalization is taken as
follows;
\begin{eqnarray}
    \label{eq:stauLR}
\left(
  \begin{array}{c}
 \tilde{\tau}_{R} \\
 \tilde{\tau}_{L}
  \end{array}
\right)
=
\left(
  \begin{array}{cc}
 c_{\tilde{\tau}} & s_{\tilde{\tau}} \\
 -s_{\tilde{\tau}}^{*} & s_{\tilde{\tau}}^{*} \\
  \end{array}
\right)
\left(
  \begin{array}{c}
 \tilde{\tau}_{1} \\
 \tilde{\tau}_{2}
  \end{array}
\right).
\end{eqnarray}
Because the mixing angle in $s_{\tilde{\tau}}$ and
$c_{\tilde{\tau}}$ is an unknown parameter, we will study the
$s_{\tilde{\tau}}$ dependence by changing it from zero to unity.

Note that the stau decay rate is approximately represented by 
\begin{eqnarray}
    \label{eq:GammaAprox}
    \Gamma(\tilde{\tau}_{1}\to\tau\tilde{F}_{1})
    &\sim&
    \frac{m_{\tilde{\tau}}}{16 \pi} \left| \delta_{1} \frac{\mu}{v}
    \right|^{2}
    \times
    \frac{1}{\Delta} \\
    &\sim&
    \left( 1 {\rm m} \times \Delta \right)^{-1}, \nonumber
\end{eqnarray}
for $F_{a} \sim 10^{10}$ GeV and $m_{\tilde{\tau}} \sim \mu \sim
{\cal O} (10^{2})$ GeV with a suppression factor $\Delta \sim
{\cal O}(10)$ caused by electroweak couplings,  mixing angles,
kinematic suppressions and so on. Thus, we can expect the decay
signals of staus within the meter-size detectors in LHC.

\section{Abundance of the flaxino DM}

The flaxino LSP abundance is determined by the thermal generation
and the non-thermal generation from the stau NLSP decay:
\begin{eqnarray}
    \label{eq:totOmega}
    \Omega_{\tilde{F_{1}}}h^{2} =
     \Omega_{\tilde{F_{1}},{\rm th}}h^{2}
    + \Omega_{\tilde{F_{1}},{\rm nonth}}h^{2}.
\end{eqnarray}
First, the freeze-out density of the relic stau can be calculated
by taking the leading $s-$wave  stau annihilation to photons whose
thermal-averaged cross section is given by $\sigma \approx
2\pi\alpha^{2}/m_{\tilde{\tau}}^{2}$ with the fine structure
constant $\alpha$.  Then, following the standard procedure
presented in ~\cite{KolbTurner}, the non-thermal density of
$\tilde{F_{1}}$ can be estimated as
\begin{eqnarray}
    \label{eq:Omega_nonth}
    \Omega_{\tilde{F_{1}},{\rm nonth}}h^{2}
    = 0.1
    \left(
      \frac{m_{\tilde{F}_{1}}}{400 \rm GeV}
    \right)
    \left(
      \frac{m_{\tilde{\tau}}}{400 \rm GeV}
    \right).
\end{eqnarray}
By using the analytical calculation  in Ref.~\cite{Chun:2000jx},
the density parameter of the thermal component of
$\tilde{F_{1}}$ is estimated to be 
\begin{eqnarray}
    \label{eq:Omega_th}
    \Omega_{\tilde{F_{1}},{\rm th}}h^{2}
  = 0.1
    \left(
      \frac{ m_{\tilde{F}_{1}} }{ m_{\tilde{q}}}
    \right)
    \left(
      \frac{\epsilon_{ \tilde{q} \tilde{q} \tilde{F}_{1} } }
      { 10^{-8}}
    \right)^{2}
    \left(
      \frac{x_{\tilde{q}}}{19.45}
    \right)^{2}
    \exp{\left[-0.98 (x_{\tilde{q}}-19.45)\right]}.
\end{eqnarray}
Here $x_{\tilde{q}} \equiv m_{\tilde{q}}/T_{\rm RH}$, with
$m_{\tilde{q}}$ the squark mass and $T_{\rm RH}$ the reheat
temperature after the lightest flaton $F_{1}$ decay,
\begin{eqnarray}
    \label{eq:TRH}
    T_{\rm RH} = 19 {\rm GeV}
    \left( \frac{m_{F_{1}}}{10^{2}{\rm GeV} }\right)^{3/2} 
    \left( \frac{F_{a}}{ 10^{10}{\rm GeV}} \right)^{-1}
    \left( \frac{B_{a}}{0.1} \right)^{-1},
\end{eqnarray}
with $B_{a}$ the branching ratio of $F_{1}$ decaying into axions.
$T_{\rm RH}$ should be sufficiently high for daughter axions produced
by the decay of $F_{1}$ not to induce $^{4}$He overproduction ( $B_{a}
< 0.1$~\cite{Chun:2000jx}) and must be also larger than
$m_{\tilde{\tau}}/25$ to thermally produce the NLSP staus.  The
quantity $\varepsilon_{\tilde{q}q\tilde{F}_1}$ denotes the
squark-quark-flaxino mixing which becomes the largest for the stop
having an order-one top Yukawa coupling $h_t$:
$\varepsilon_{\tilde{q}q\tilde{F}_1} = h_t N_{04}$ with $N_{04} \simeq
\delta_{1} c_{\beta} \sim {\cal O}(v/F_{a})$.

\section{The flaton and flatino mass spectrum}
By assuming the superpotential given in~(\ref{eq:superpot}), we have a
scalar potential including soft supersymmetry breaking terms,
\begin{eqnarray}
    \label{eq:Vsoft}
    V_{\rm soft} = f \frac{A_{f}}{M_{p}} P^{3} Q + h.c.
\end{eqnarray}
Then the flaton
masses are given by
\begin{eqnarray}
    \label{eq:mF12}
    m_{F_{1,2}} &=& \frac{\tilde{\mu}}{\sqrt{2}}  
    \sqrt{3 (12-\xi) + x^{2} (12+\xi) \pm |12 -\xi| \sqrt{x^{4} + 42
    x^{2} +9} },
    \\
    \label{eq:mFdash}
    m_{F'} &=& \tilde{\mu} \sqrt{\xi(x^{2} +9)},
\end{eqnarray}
where $\tilde{\mu}\equiv \mu f/h$ and  $\xi \equiv -A_{f}/(f \mu_{0})$
with $\mu_{0} = \langle P \rangle \langle Q \rangle /2M_{p}$ Then, the
flatino masses are represented by
\begin{eqnarray}
    \label{eq:mtildeF}
    m_{\tilde{F}_{1,2}} = 3 \tilde{\mu} \left( \sqrt{x^{2}+1 } \pm 1\right).
\end{eqnarray}
Allowed parameter regions for $x$ and $\xi$ of our interest (i.e,
parameters where the flatons cannot decay into flatinos,
$m_{F_{1,2},F'} < 2 m_{\tilde{F}_{1,2}}$) are given  in (28) in
Ref.~\cite{Chun:2000jx}.

\section{Results}

We present our results on the DM abundance and the stau lifetime by
taking some typical parameter sets of the flat DFSZ model.  Following
the notations and the parameters presented in the previous sections
and Ref.~\cite{Chun:2000jx}, we take a set of $x=4$, $\xi = 13$, $f/h
= 1/24$, $A_{h}/\mu = -1$, $A_{f}/\mu = 13/24$, $\tan\beta = 3$, and
$M_{2} = 2 M_{1}$. In this paper we show two cases of the mass
parameters 1) $\mu$ = 287 GeV ($\tilde{\mu} = 12$) and $M_{1} = 130$
GeV, and 2) $\mu$ = 813 GeV ($\tilde{\mu}=34$) and $M_{1} = 530$ GeV.

We can calculate the mass spectrum in GeV unit of the flatons
$F_{1,2}$, the pseudoscalar flaton $F'$, and the flatinos
$\tilde{F}_{1,2}$ which is given as
\begin{eqnarray}
    \label{eq:mspectrum}
    \left[ m_{F'}, m_{F_{2}},  m_{F_{1}},  m_{\tilde{F}_{2}},
    m_{\tilde{F}_{1}}  \right]  
  =
\left\{
  \begin{array}{l}
      \left[ 216, 175, 162, 184, 112 \right] 
      \quad \left(\tilde{\mu}= 12 \right),
      \\
      \left[ 616, 500, 462, 525, 320 \right] 
      \quad \left(\tilde{\mu}= 34 \right).
  \end{array}
\right.
\end{eqnarray}

In each case, by using Eq.(\ref{eq:TRH}) the reheat temperature after
the thermal inflation (from the decay of the lightest flaton $F_{1}$)
is given by $T_{\rm RH}=39$ ( $T_{\rm RH}=189$ GeV) for $\tilde{\mu} =
12$ ($\tilde{\mu} = 34$).

In addition, we can diagonalize the 6$\times$6 mass matrix by using
the approximated 5$\times$5 method discussed in
Sec.~\ref{sec:lifetime} and get the mass spectrum of four neutralinos
plus two flatinos. Then we can calculate the mass of the lightest
neutralino $\tilde{\chi}_{0}$ (or $\tilde{N}_{1}$) and get 1)
$m_{\tilde{\chi}_{0}} =$ 120 GeV and 2) 525 GeV, respectively.

\begin{figure}
    \begin{center}
    \includegraphics[width=100mm,clip,keepaspectratio]{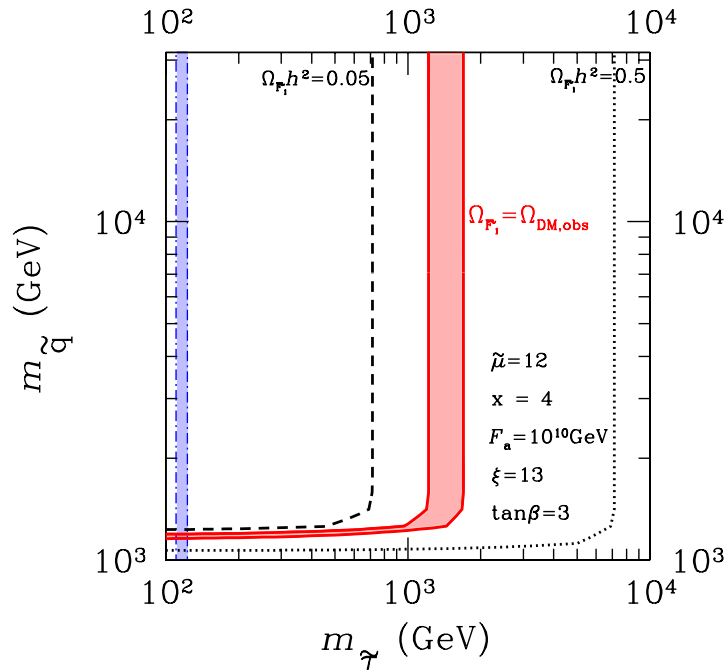}
        \caption{Allowed regions of flaxino  dark matter. 
        The axes correspond to the masses of the stau and the squark
        (stop).  The other parameters are as described in the text,
        with the choices $\tilde{\mu}$=12, $x$=4, $\xi$=13,
        $\tan\beta$=3, and $F_{a} = 10^{10}$ GeV.  The region between
        the two  solid lines satisfies  the WMAP constraints on
        $\Omega_{\rm LSP}$ given in Eq.~(\ref{eq:OmegaWMAP3}).  The
        lightest flatino LSP  and the stau NLSP  can be realized in
        the  region between the two dot-dashed lines.  The
        intersection of these regions is allowed. }
      \label{fig:omega_mu12}
    \end{center}
\end{figure}

\begin{figure}
    \begin{center}
    \includegraphics[width=100mm,clip,keepaspectratio]{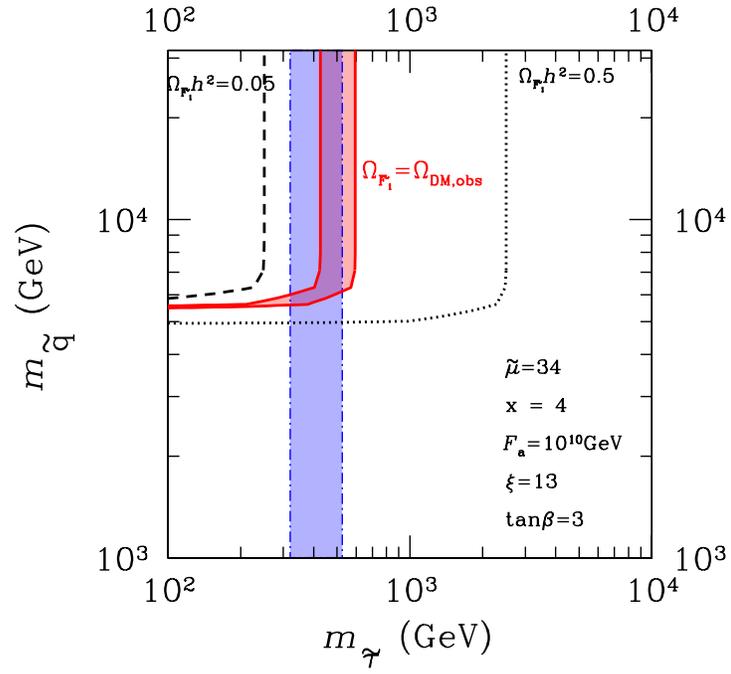}
        \caption{
        Same as Fig.~\ref{fig:omega_mu12} but for $\tilde{\mu}$=34.
        }
        \label{fig:omega_mu34}
    \end{center}
\end{figure}

In Figs.~\ref{fig:omega_mu12} and \ref{fig:omega_mu34} we plot the
observationally-allowed region which is enclosed by two solid lines in
($m_{\tilde{\tau}}$,$m_{\tilde{q}}$) plane for $\tilde{\mu}$=12 and
34, respectively. Here we have  $F_{a} = 10^{10}$ GeV. In both
figures, the vertical area enclosed by solid lines comes from the
non-thermal contribution and the narrow horizontal area from the
thermal contribution which is very sensitive function of
$x_{\tilde{q}}$.  The vertical band between two dot-dashed lines
denotes a region where the lightest flatino (flaxino) LSP  and the
stau NLSP are realized. In Fig.~\ref{fig:omega_mu34}, we see that the
flaxino dark matter naturally agrees with observations in broad
parameter spaces of $m_{\tilde{q}}$ at the weak scale.

\begin{figure}
    \begin{center}
\includegraphics[width=100mm,clip,keepaspectratio]{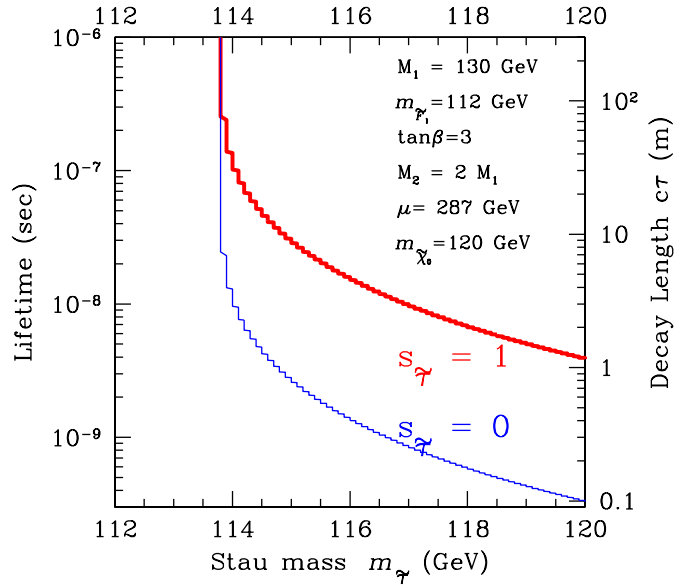}
        \caption{
        Stau lifetime decaying into the lightest flatino. The thick
        (thin) solid line denotes the case of the mixing angle
        $s_{\tilde{\tau}} = 1$ ($s_{\tilde{\tau}} = 0$). Here we have
        adopted $F_{a} = 10^{10}$GeV and the electroweak scale $v$ =
        264 GeV. Note that $m_{\tilde{F}_{1}}$ =112 GeV, and the mass
        of the lightest neutralino $\tilde{N}_{1}$
        ($m_{\tilde{\chi}_{0}}=m_{\tilde{N}_{1}} = 120$ GeV) is
        smaller than that of $\tilde{F}_{2}$ ($m_{\tilde{F}_{2}}$ =184
        GeV) in this case.}
        \label{fig:lifetime}
    \end{center}
\end{figure}

\begin{figure}
    \begin{center}
\includegraphics[width=100mm,clip,keepaspectratio]{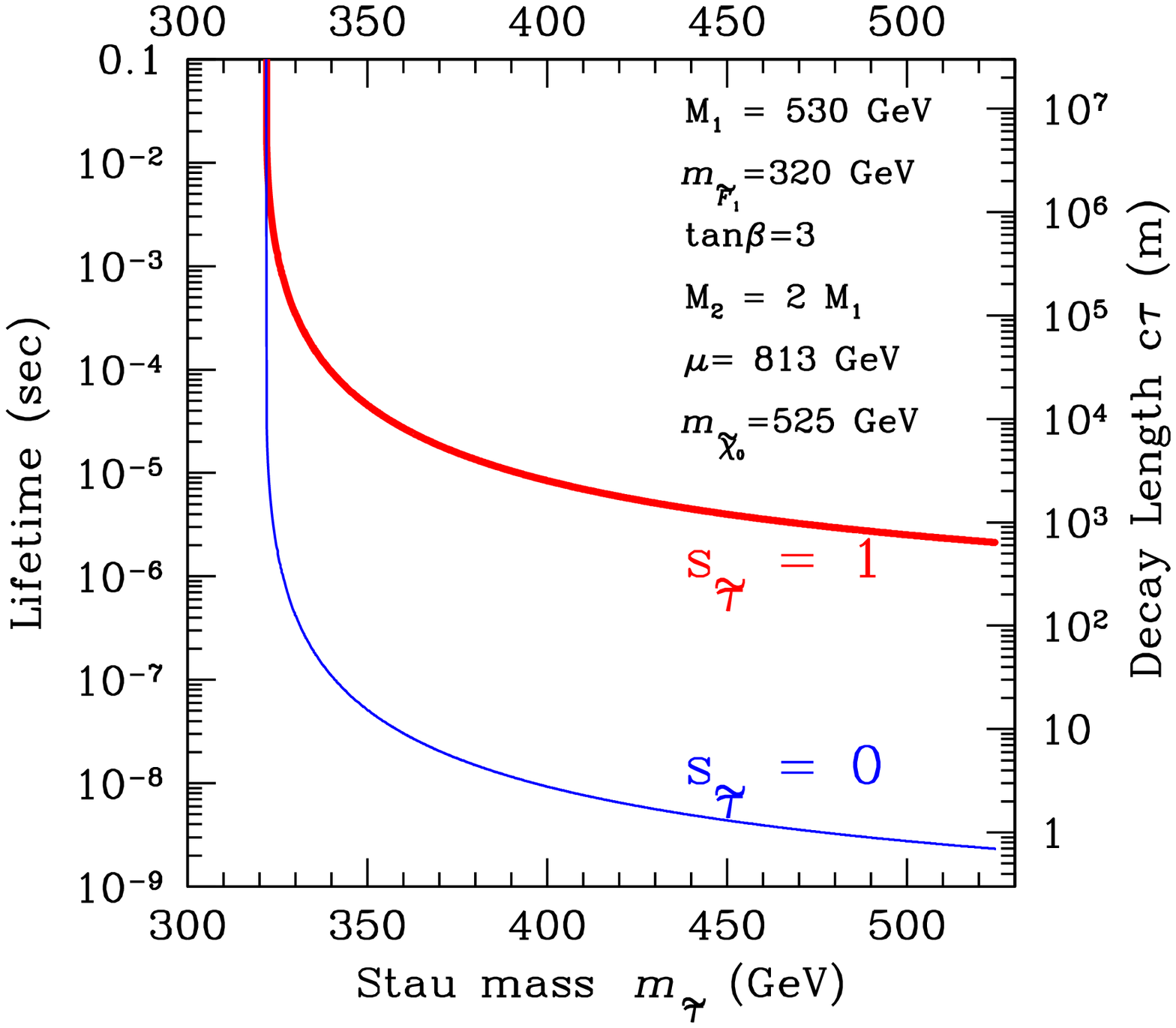}
        \caption{
        Same as Fig.\ref{fig:lifetime} but for $\tilde{\mu}$=34
        ($\mu=813$ GeV) and $M_{1} = 530$ GeV. Then,
        $m_{\tilde{F}_{1}} =$ 320 GeV, and  $m_{\tilde{N}_{1}} \simeq
        m_{\tilde{F}_{2}} =$ 525 GeV.}
        \label{fig:lifetime2}
    \end{center}
\end{figure}

In Figs.~\ref{fig:lifetime} and~\ref{fig:lifetime2} we plot the
lifetime of stau as a function of stau mass in case of
$\tilde{\mu}=12$, and 34, respectively. The thick (thin) solid line
denotes the case of the mixing angle $s_{\tilde{\tau}} = 1 (0)$. Here
we have taken $F_{a} = 10^{10}$ GeV to minimize the decay length. In
case of $\tilde{\mu}=12$, the decay length is shown to be around 1 m
and thus the stau NLSP decay to the flaxino LSP will be possibly
searched for at LHC.  Note that the lifetime scales $\propto
F_{a}^{2}$. From these figures we see that the lifetime is
sufficiently shorter than $10^{3}$ sec excluding the special case of
degenerate masses $m_{\tilde{\tau}} - m_{\tilde{F}_{1}} \sim m_{\tau}$
($\sim 1.8$ GeV). Hence this scenario is not constrained by the
$^{6}$Li overproduction through BBN if we adopt the standard range of
$F_{a} = 10^{10}$ GeV -- $10^{12}$ GeV.  In other words, even lager
value $F_{a} \sim 10^{14}$ GeV can be allowed from the viewpoint of
BBN. Such a high value might not lead to the overclosure of the
universe by the axions because of a sufficiently low reheating
temperature and large entropy productions after the thermal
inflation~\cite{CCK}. See also Ref.\cite{Grin:2007yg} for  recent
detailed analyses of the thermally-produced axions for a relatively
low reheating temperature.

\section{Conclusions}

The lightest supersymmetric particle (LSP) is a dark matter candidate.
It is usually said that the LSP will be either the lightest
superpartner of a Standard Model particle (so called LOSP), the
gravitino or the axino.  For cosmological constraints and collider
signals of the LSP, it matters what is the NLSP.  In this paper we
revisited the axino-like LSP arising in the flat DFSZ axion models for
the case of the stau NLSP.

Following \cite{Chun:2000jx}, we assumed that the spontaneous breaking of
Peccei-Quinn symmetry
comes only from supersymmetry breaking. Then all of the PQ particles
except the axion will have masses of order the gravitino mass.
The scalar particles are called flatons and their superpartners are called
flatinos.

In this scenario the saxion is a linear combination of flatons and the 
axino a linear combination of flatinos, but they are not mass eigenstates and
have no special status. So the LSP candidate from the PQ sector is not the 
axino, but the lightest flatino which we have dubbed the flaxino.
We have explored the simplest version of this scenario,
which  has two flatinos and assumes  gravity 
mediated supersymmetry breaking. 

We  have found that the stau NLSP decay to the flaxino LSP is fast
enough to maintain the standard predictions of Big Bang
Nucleosynthesis.  The big difference from the gravitino dark matter
scenario is that the flaxino/axino coupling ($\sim 1/F_a$) is much
larger than the gravitino coupling ($\sim 1/M_P$), which makes the
stau NLSP decay much faster.  In the DFSZ axion model,  the
stau-flaxino/axino coupling appears at tree level as a consequence of
the $\mu$ term mixing between the higgsino and flaxino/axino.  This
makes the NLSP decay more efficient compared with the KSVZ axion
models.

In a specific DFSZ model whose parameters are constrained by various
cosmological considerations, we identified the region of the stop and
stau masses which is compatible with the required dark matter density.
Within the conventional bound on the axion scale $F_a \lesssim
10^{12}$ GeV, the stau decay to the flaxino LSP never becomes
dangerous for BBN. Furthermore, with the chosen parameter sets in the
first case ($\tilde{\mu}=12$), the stau decay length turns out to be
${\cal O}$(1) m so that a fair number of stau decays can be captured
in colliders when the axion scale is close to its lower bound, $F_a
\sim 10^{10}$ GeV. In the second case ($\tilde{\mu}=34$), the flaxino
LSP is naturally realized in the broad parameter space of the squark
mass. Although the decay length is longer than ${\cal O}$(10) m, the
additional detectors proposed by \cite{Hamaguchi:2004df,Feng:2004yi,De
Roeck:2005bw} will catch the staus even in this case.

Let us finally remark that, in our scheme, the unstable gravitino
decaying to ordinary superparticle or flaxino does not contradict with
BBN as the primordial gravitino abundance will be diluted away by the
entropy production after thermal inflation caused by the flat
direction in the model.  A gravitino even lighter than the flaxino
is also allowed since the flaxino decay to the gravitino and the axion
causes no trouble with BBN.

\section*{Acknowledgements}
This work was supported by the Science Research Center Program of the
Korea Science and Engineering Foundation through the Center for
Quantum Spacetime(CQUeST) of Sogang University with grant number
R11--2005--021 (H.B.K.). The research at Lancaster is supported by
PPARC grant PP/D000394/1 and by EU grants MRTN-CT-2004-503369 and
MRTN-CT-2006-035863, the European Union through the Marie Curie
Research and Training Network "UniverseNet".



\end{document}